\documentclass[twocolumn,graphicx,pra]{revtex4}
\usepackage{amssymb}
\usepackage{epsfig}
\usepackage{bm}
\usepackage{amsmath}
\usepackage{graphicx}
\usepackage{epstopdf}

\begin{document}

\title{Quantum-orbit analysis of above threshold ionization by intense spatially inhomogeneous field}
\author{T. Shaaran$^{1,2}$, M. F. Ciappina$^{1,3}$ and M. Lewenstein$^{1,4}$}
\affiliation{$^1$ICFO-Institut de Ci\`{e}nces Fot\`{o}niques, 08860 Castelldefels,
Barcelona, Spain\\
$^2$CEA-Saclay, IRAMIS, Service des Photons, Atomes et Molecules, 91191
Gift-sur-Yvette, France\\
$^3$Department of Physics, Auburn University, Auburn, Alabama, 36849, USA\\
$^4$ICREA-Instituci\'o Catalana de Recerca i Estudis Avan\c{c}ats, Lluis Companys 23,
08010 Barcelona, Spain}


\date{\today}

\begin{abstract}
We perform a detailed analysis of above threshold ionization (ATI) in atoms
within the strong field approximation (SFA) by considering spatially
inhomogeneous monochromatic laser fields. The locally enhanced field induced by resonance plasmons is an example for such inhomogeneous fields.
We investigate how the individual pairs of quantum orbits contribute to the photoelectron spectra and the
angular electron momentum distributions. We demonstrate that the quantum
orbits have a very different behavior in the spatially inhomogeneous field when compared to the homogeneous
field. In the case of inhomogeneous fields, the ionization and rescattering
times differ between neighboring cycles, despite the field being monochromatic. Indeed, the contributions from one cycle may lead to a
lower cutoff, while another may develop a higher cutoff. Within our
model, we show that the ATI cutoff extends far beyond the semiclassical
cutoff, as a function of inhomogeneity strength. Furthermore, the angular
momentum distributions have very different features compared to the
homogeneous case. For the neighboring cycles, the electron momentum 
distributions do not share the same absolute momentum, and they do not have the
same yield.
\end{abstract}

\maketitle

\section{INTRODUCTION}

\label{introduction}

In the context of the interaction of matter with strong laser fields, above threshold ionization (ATI) ~\cite{milosevic_rev,Agostini1979}
has attracted considerable interest, since the behavior of the laser-ionized
electrons during ATI serve as a very valuable tool for laser pulse
description. The electric field in a few-cycle pulse can be characterized by
its duration and by the so-called Carrier Envelope Phase (CEP)~\cite{Wittmann2009,kling2008}. 
In particular, for a better control of the system
on an attosecond time-scale it is important to find reliable and direct
schemes to measure the absolute phase of few-cycle pulses. The investigation
of ATI generated by few-cycle driving laser pulses plays a key role in the
CEP characterization due to the sensitivity of the energy and angle-resolved
photoelectron spectra to the value of the laser electric field absolute
phase~\cite{paulus_cleo,sayler}. In addition, it seems that the high energy
region of the photoelectron spectra is most sensitive to the absolute CEP
and consequently ATI yields with larger cutoffs is needed in order to describe it~%
\cite{milosevic_rev,paulus2003}.

Recent studies have demonstrated that the cutoff of the high harmonic
generation and photoelectron spectra could be extended further by employing
spatially inhomogeneous fields~\cite{tahirrapid,joseprl2013,ATICiappina2012,HHGSFAShaaran2012,HHGCiappina2012,Yavuz2012,Kim2008,Kimnew,husakou}. 
In particular, recent experiments using a combination of rare gases and
plasmonic nanostructures have demonstrated that the harmonic cutoff of the
gases could be extended further by using the locally enhanced field
induced by coupling laser pulses to nanosystems\cite{Kim2008,Kimnew}. 
In such system, the locally enhanced field is not spatially homogeneous
in the region of electron dynamics. This is a result of the strong
confinement of the plasmonics spots and the distortion of the electric field
by the surface plasmons induced by the nanosystem. In addition, recently,
solid state nanostructures have been employed as a target to study the
photoelectron emission by few intense laser pulse~\cite{klin,peternature}.
In this process, the emitted electrons have energy far beyond the usual
cutoff for noble gases (see e.g.~\cite{peternature,peterprl2006,peterprl2010,peterjpbreview,jensdombi,ropers}).

Hitherto, the theoretical approaches, including solving the Time Dependent
Schr\"{o}dinger Equation (TDSE) and the Strong Field Approximation (SFA),
for studying strong field phenomena are mainly based on the assumption that
the laser electric field is spatially homogeneous in the region of electron dynamics~\cite{Protopapas1997,Brabec2000}. This, hypothesis, however, is no longer valid for the locally enhanced
plasmonic field. Indeed, in such system the strong confinement of the
electrons in the plasmonic hot spots generates a spatially inhomogeneous
inhomogeneous electric field. It means the driven electron will experience a Lorentz force which depends on position and its subsequent motion will be strongly affected. As a result, new physics will emerge in the interaction between matter and strong laser fields. Until now, few theoretical approaches
have been developed investigating the strong field phenomena in such
spatially inhomogeneous fields~\cite{ATICiappina2012,HHGSFAShaaran2012,HHGCiappina2012,Yavuz2012}.

In the present paper, we employed SFA to investigate the ATI by resonant
plasmon field enhancement. In addition, we use a SFA based on saddle point
methods rather than a full numerical SFA approach to obtain the ATI
photoelectron spectra and electron momentum distributions. The saddle point
methods lead to equations that can be directly related to the classical
equations of motion of an electron in a laser field. As a result, they
provide a space-time picture which gives us additional physical insight. In
this work, which based on this method, we investigate the individual
electron trajectories and demonstrate their contributions to the ATI spectra
and the electron momentum distributions. In addition, since the imaginary
part of the saddle point equations can be related to the width of the
potential barrier through which the electron tunnels, we examine the
ionization probability of the electron for each trajectory.

This article is organized as follows. In Sec.~\ref{theory}, we present the
SFA transition amplitudes for both direct and rescattered ATI driving by
lineally spatially inhomogeneous field. We start from common expressions
based on homogeneous fields(Sec.~\ref{transitionamp}), and subsequently
show the modified transition amplitude by considering inhomogeneous fields with linear dependence
(Sec.~\ref{SFNEQ}). In Sec.~\ref{gmorbits}, we discuss the saddle-point
equations and analyze them in terms of quantum orbits in parallel to their
classical counterpart electron trajectories, in terms of both electron
energy and momentum . In the next section, Sec.\ref{spectrasec}, we present the ATI
spectra based on the analysis given in Sec \ref{gmorbitsenergy}. In Sec.~\ref{momentumDist},
we provide angular momentum distributions of the direct and
rescattered ATI based on the analysis given in Sec.~\ref{gmorbitsmomentum}.
Finally, in Sec.~\ref{conclusion} we conclude the paper with a few
summarizing remarks.

\section{Theory}

\label{theory}

\subsection{Transition Amplitude}

\label{transitionamp}

The strong field approximation (SFA) is based on two assumptions, namely (i)
the influence of the laser field is neglected when the electrons are bound
to their target atoms and (ii) the binding ionic potential is neglected when
the electrons are in the continuum. As a result, the free electrons in the
continuum are described by field-dressed plane waves, known as
Volkov states~\cite{Gordon1926,Volkov1935}. Based on the assumption that the
laser electric field does not change with respect of the position in the
region where the electron dynamic takes place, the well-established Keldysh-Faisal-Reiss (KFR) model~\cite{kfr1,kfr2,kfr3,kfr4}, gives the SFA transition amplitude for
the direct and rescattered ATI as (in atomic units):
\begin{equation}
M_{dir}=-i\int_{-\infty }^{\infty }\hspace*{-0.1cm}dt^{\prime }\hspace*{0.1cm%
}V_{p0}e^{iS_{0}(t^{\prime })}  \label{TranDir}
\end{equation}
and
\begin{equation}
M_{res}=-\int_{-\infty }^{\infty }\hspace*{-0.2cm}dt\hspace*{-0.1cm}%
\int_{-\infty }^{t}\hspace*{-0.1cm}dt^{\prime }\hspace*{-0.0cm}\int
d^{3}kV_{pk}V_{k0}e^{iS_{0}(\mathbf{k},t,t^{\prime })},  \label{TranResc}
\end{equation}
where the corresponding actions are given by
\begin{equation}
S_{0}(t^{\prime })=-\int_{t^{^{\prime }}}^{\infty }\hspace{-0.1cm}\frac{[%
\mathbf{p}+\mathbf{A}(\tau )]^{2}}{2}d\tau +I_{p}t^{^{\prime }}
\label{ActionDir}
\end{equation}
and
\begin{equation}
S_{0}(\mathbf{k},t,t^{\prime })=-\int_{t}^{\infty }\hspace{-0.1cm}\frac{[%
\mathbf{p}+\mathbf{A}(\tau )]^{2}}{2}d\tau -\int_{t^{\prime }}^{t}\hspace{%
-0.1cm}\frac{[\mathbf{k}+\mathbf{A}(\tau )]^{2}}{2}d\tau +I_{p}t^{^{\prime }}.
\label{ActionResc}
\end{equation}
Thereby, $\mathbf{k}$, $\mathbf{p}$ and $I_{p}$ denote the
intermediate and final momentum of the electron and the ionization potential of the field-free bound state $\left\vert \phi _{0}\right\rangle $, respectively

The prefactors of the transition amplitudes $M_{dir}$ and $M_{res}$ read
\begin{eqnarray}
V_{\mathbf{p}0} &=&\langle \widetilde{\mathbf{p}}(t^{\prime })|V|\phi
_{0}\rangle  \notag \\
&=&\frac{1}{(2\pi )^{3/2}}\int d^{3}\mathbf{r}\exp [-i\widetilde{\mathbf{p}}%
(t^{\prime })\cdot\mathbf{r}]V(\mathbf{r})\phi _{0}(\mathbf{r}),  \label{prefVp0}
\end{eqnarray}

\begin{eqnarray}
V_{\mathbf{k}0} &=&\langle \widetilde{\mathbf{k}}(t^{\prime })|V|\phi
_{0}\rangle  \notag \\
&=&\frac{1}{(2\pi )^{3/2}}\int d^{3}\mathbf{r}\exp [-i\widetilde{\mathbf{k}}%
(t^{\prime })\cdot\mathbf{r}]V(\mathbf{r})\phi _{0}(\mathbf{r})  \label{prefVk0}
\end{eqnarray}
and
\begin{eqnarray}
V_{\mathbf{pk}} &=&\langle \widetilde{\mathbf{p}}(t)|V|\widetilde{\mathbf{k}}%
(t)\rangle  \notag \\
&=&\frac{1}{(2\pi )^{3}}\int d^{3}\mathbf{r}\exp [-i(\widetilde{\mathbf{p}}-%
\widetilde{\mathbf{k}})\cdot\mathbf{r}]V(\mathbf{r})\phi _{0}(\mathbf{r})
\label{prefVpk}
\end{eqnarray}
where $V$ gives the interaction of the system with the laser field. Here, 
$\widetilde{\mathbf{k}}(t^{\prime })=\mathbf{k}$ and $\widetilde{\mathbf{p}}%
(t^{\prime })=\mathbf{p}$ in the velocity gauge, and $\widetilde{\mathbf{k}}%
(t^{\prime })=\mathbf{k+A(t,)}$ and $\widetilde{\mathbf{p}}(t^{\prime })=%
\mathbf{p+A(t}^{\prime }\mathbf{)}$ in the length gauge.

\subsection{Strong Field and Newton's Equations}

\label{SFNEQ}

We now study a case in which the laser field has a spatially
inhomogeneous character in the region where the electron motion takes place.
We start by examining how the action of the SFA is connected to the
classical electron trajectories. In the SFA, the action is defined in terms
of the vector potential field, which is the counterpart of the velocity $\dot{x}(t)$ in Newton's equation of motion. The laser potential $V_{L}$
due to the laser field $E(t,x)$ is defined as
\begin{equation}
V_{L}=xE(t,x),  \label{potfieldhm}
\end{equation}
and the Newtonian equation of motion for an electron in this field is given by
\begin{equation}
\ddot{x}(t)=\nabla _{x}V_{L}=-x\nabla _{x}E(t,x)-E(t,x).  \label{Newton}
\end{equation}

For the homogeneous case, in which the laser field does not have spatially
dependency and the laser electric field is just $E(t)$, the Newton equation
reads $\ddot{x}(t)=-E(t)$. On the other hand, if the spatial dependence of the enhanced laser electric field is
perturbative and linear with respect to position, then the field can be approximated as
\begin{equation}
E(t,x)\simeq E(t)(1+\epsilon x),  \label{nhmfield}
\end{equation}%
where $\epsilon \ll 1$ is a parameter that characterize the strength of the
inhomogeneity.

Indeed, the above approximation corresponds to the first term of the actual
field of a plasmonic nanostructure with spherical shape~\cite{Klingnano2011}. 
By substituting Eq.~(\ref{nhmfield}) into (\ref{Newton}), we have
\begin{equation}
\ddot{x}(t)=-E(t)(1+2\epsilon x(t)).  \label{Newtonh1}
\end{equation}%
This is the effective laser electric field that the electron feels along the
trajectory $x(t)$, which describes its motion once it is laser freed to the continuum.

Classically, the electron trajectory can be found by solving Eq.~(\ref{Newtonh1}). In here, we solve it by applying the Picard iteration~\cite{EdwardG1999} method and restrict ourselves to the first order (for more
details see~\cite{HHGCiappina2012} ). Based on the condition that the
electron starts its movement at the origin with zero velocity, i.e. $x(0)=0$
and $\dot{x}(0)=v(0)=0$, we obtain
\begin{equation}
x(t)=\beta (t)-\beta (t_{0})-A(t_{0})(t-t_{0}).  \label{position}
\end{equation}
with $\beta (t)=\int_{0}^{t}dt^{\prime }A(t^{\prime })$. In
addition we assume that at time $t_{0}$ the potential field is zero, then
\begin{equation}
x(t)=\int^{t}dt^{\prime }A(t^{\prime }).  \label{newposition}
\end{equation}

By using Eqs.~(\ref{Newtonh1})and (\ref{newposition}) the effective
vector potential along the electron trajectory $A_{tr}(t,x)$ reads
\begin{equation}
A_{tr}(t)=A(t)+2\epsilon A_{c}(t),  \label{newpotentialfield}
\end{equation}
where
\begin{equation}
A_{c}(t)=\int^{t}dt^{\prime \prime }A(t^{\prime \prime })-\int^{t}dt^{\prime
\prime }A^{2}(t^{\prime \prime }),
\end{equation}
and
\begin{equation}
A(t)=-\int_{\infty }^{t}dt^{\prime }E(t^{\prime }).
\end{equation}

The next step is to incorporate the above defined inhomogeneous vector potential field into the general expression of the transition
amplitudes, i.e. Eqs.~(\ref{TranDir}) and (\ref{TranResc}) for the direct and
rescattered ATI, respectively. Consequently, we replace the vector potential $\mathbf{A}(t)$ in $M_{dir}$ and $M_{res}$
by the one defined in Eq.~(\ref{newpotentialfield}), respectively. As a result, the modified
actions yield
\begin{eqnarray}
S(t^{\prime }) &=&S_{0}(t^{\prime })-2\epsilon \int_{t^{^{\prime }}}^{\infty
}\hspace{-0.1cm}A_{c}(\tau )[\mathbf{p}+\mathbf{A}(\tau )]d\tau  \notag \\
&&-2\epsilon ^{2}\int_{t^{^{\prime }}}^{\infty }\hspace{-0.1cm}%
A_{c}^{2}(\tau )d\tau  \label{NewActionDir}
\end{eqnarray}
and
\begin{eqnarray}
S(\mathbf{k},t,t^{\prime }) &=&S_{0}(\mathbf{k},t,t^{\prime })-2\epsilon
\int_{t}^{\infty }\hspace{-0.1cm}A_{c}(\tau )[\mathbf{p}+\mathbf{A}(\tau
)]d\tau  \notag \\
&&\hspace{-0.9cm}-2\epsilon \int_{t^{^{\prime }}}^{t}\hspace{-0.2cm}%
A_{c}(\tau )[\mathbf{k}+\mathbf{A}(\tau )]d\tau -2\epsilon ^{2}\hspace{-0.2cm%
}\int_{t^{^{\prime }}}^{\infty }\hspace{-0.2cm}A_{c}^{2}(\tau )d\tau
\label{NewActionResc}
\end{eqnarray}
where $S_{0}(t^{\prime })$ and $S_{0}(\mathbf{k},t,t^{\prime })$ are defined
in Eqs.~(\ref{ActionDir}) and (\ref{ActionResc}), respectively.

\subsection{Saddle-point equations}

\label{saddleEqs}

For large driving-field intensities, the transition amplitudes Eqs.~(\ref{TranDir}) and (\ref{TranResc}) are strongly oscillatory integrals which can be
evaluated using the saddle-point or steepest descent method~\cite{Bleinstein1986,Salieres2001}.
The solutions of the saddle point equations are directly related to the
classical trajectories, which allow us to compare them with the quantum orbits. This method requires obtaining the saddle points where
the action Eq.~(\ref{ActionDir}) for the direct ATI and the action Eq.~(\ref{ActionResc})
for the rescattered ATI are stationary, i.e. for which $%
\partial _{t}S(t^{\prime })=\partial _{t}S(\mathbf{k},t,t^{\prime
})=\partial _{t^{\prime }}S(\mathbf{k},t,t^{\prime })=\partial _{k}S(\mathbf{%
k},t,t^{\prime })=0$. In this paper, we use a specified steepest descent
method called uniform approximation to take care of those saddles points
which are not well separated (for a detailed discussion see Ref.~\cite%
{Faria2002}). For the direct ATI, the stationary condition upon $t^{\prime
} $ lead to the saddle-point equation
\begin{equation}
\left[ \mathbf{p}+\mathbf{A}(t^{\prime })\right] ^{2}+4\epsilon \lambda
(t^{\prime })=-2I_{p}  \label{saddle1dir}
\end{equation}
with
\begin{equation}
\lambda (t^{\prime })=A_{c}(t^{\prime })[\mathbf{p}+\mathbf{A}(t^{\prime
})]+\epsilon A_{c}^{2}(t^{\prime }).  \label{saddle4dir}
\end{equation}

For the rescattered ATI, the stationary conditions upon $t,t^{\prime }$ and $%
k$ lead to the saddle-point equations
\begin{equation}
\left[ \mathbf{k}+\mathbf{A}(t^{\prime })\right] ^{2}+4\epsilon \kappa
(t^{\prime })=-2I_{p}  \label{saddle1}
\end{equation}%
\begin{equation}
\int_{t^{\prime }}^{t}d\tau \lbrack \mathbf{k}+\mathbf{A}(\tau )]+4\epsilon
\eta \mathbf{(\tau )}=0  \label{saddle2}
\end{equation}%
and
\begin{equation}
\frac{\left[ \mathbf{p}+\mathbf{A}(t)\right] ^{2}}{2}=\frac{\left[ \mathbf{k}%
+\mathbf{A}(t)\right] ^{2}}{2}+2\epsilon A_{c}(t)[\mathbf{k}-\mathbf{p}]
\label{saddle3}
\end{equation}%
with
\begin{equation}
\kappa (t^{\prime })=A_{c}(t^{\prime })[\mathbf{k}+\mathbf{A}(t^{\prime
})]+\epsilon A_{c}^{2}(t^{\prime }),  \label{saddle4}
\end{equation}%
and
\begin{equation}
\eta (\tau )=\int_{t^{^{\prime }}}^{t}\hspace{-0.1cm}A_{c}(\tau )d\tau .
\label{saddle5}
\end{equation}%
Equations (\ref{saddle1}) and (\ref{saddle1dir}) express the conservation law
of energy for the electron tunnel ionized at the time $t^{\prime }$. Furthermore, Eq.~(\ref{saddle2}) constrains the intermediate momentum of the electron and guarantees that the electron returns to its parent ion. Finally, Eq.~(\ref{saddle3}) gives the energy conservation of the electron at the time $t$,
when upon its return inelastically rescatters with the core.

The inhomogeneous character of the laser field gives the additional terms $\lambda (t^{\prime})$, $\kappa (t^{\prime})$
and $\eta (t)$ in the Eqs.~(\ref{saddle1dir}), (\ref{saddle1}), and
(\ref{saddle2}), respectively. These terms vanish for the homogeneous case, i.e when $\epsilon
=0$. Furthermore, for the homogeneous case Eq.~(\ref{saddle1}) has no real solutions, unless
$I_{p}\rightarrow 0$, due to tunnel ionization, which has no classical counterpart. Therefore, the solutions of the saddle equations are generally complex. For the inhomogeneous case, however, it is not very upfront to constrain the limits, in which the solutions of Eq.~(\ref{saddle1}) are real. 
Nevertheless, in our case, the electron will most likely reach the continuum via tunnel ionization,
since $\epsilon$ is a very small parameter. It means that the solutions of these saddle point equations are still expected to be
complex. In addition, the maximum kinetic energy that the electron gains in
the continuum now depends on the nonhomogeneous character of the field (i.e. on the value of $\epsilon$) and $A_{c}(t)$, and it does not have anymore the conventional value of $3.17U_{p}$, where $U_{p}=E_{0}^{2}/(4\omega ^{2})$ is the ponderomotive energy~\cite{Corkum1993,sfa}.

\section{RESULTS}

\label{results}

\subsection{Quantum orbits}

\label{gmorbits}

\begin{figure}[h]
\begin{center}
\includegraphics[width=9cm]{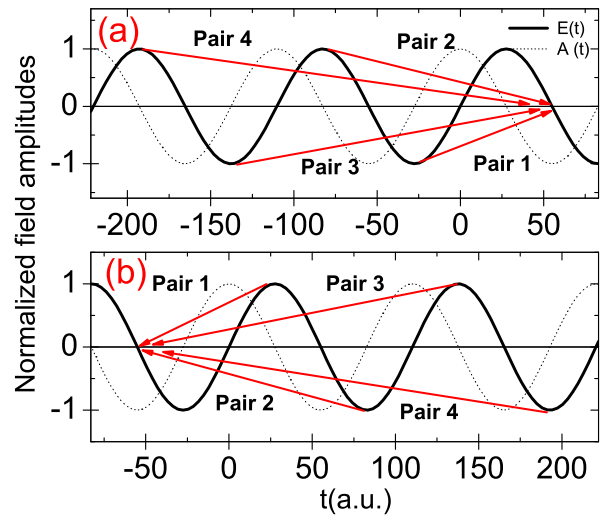}
\end{center}
\caption{(color online) Schematic representation of the laser electric field $\mathbf{E}(t)$
and the corresponding vector potential field $\mathbf{A}(t)$ for a monochromatic
field defined by $\mathbf{E}(t)=E_0 \sin(\protect\omega t) e_{x}$ (we show only 3
cycles) vs. $t$ in a.u. The arrow indicates the classical times around which the electrons
leave to the continuum and return to the core (approximately the field maxima and the field crossings, respectively). The
pairs of the orbits are indicated by the labels $Pair_{n}$, where $n$ ranges
from $1$ to $4$. The fields are normalized to $\mathbf{A}(t)/A_{0}$ and $%
\mathbf{E}(t)/E_{0}$ , where $A_{0}$ and $E_{0}$ are the field peak amplitudes.}
\label{pulse}
\end{figure}

In this section, we study the solutions of the saddle-point equations to
examine the role of individual trajectories in the ATI photoelectron spectra cutoff and angular momentum distributions by performing
a quantum-orbit analysis of the problem. The concept of the quantum-orbits emerges from the fact that the solutions of
the saddle-point equations are related to the classical trajectories of the
electron in the laser field and, in addition, provides information on quantum aspects such as
tunneling and interference. To get a better insight into the inhomogeneous
case, we employ a monochromatic field:
\begin{equation}
\label{monochr} 
E(t)=E_{0}\sin (\omega t)e_{z},
\end{equation}
where $e_{z}$ is the polarization vector along the $z$-axis (see Fig.~\ref{pulse}).

By using the relationship defined in Eq.~(\ref{newpotentialfield}) and
applying some trigonometric identities, the laser effective vector potential field associated with the monochromatic field Eq.~ (\ref{monochr}) along the electron trajectory reads
\begin{equation}
A_{tr}(t)=A_{0}\cos (\omega t)+2\epsilon A_{c}(t),
\label{monochpotentialfield}
\end{equation}
where $A_{0}=E_{0}/\omega $ and
\begin{equation}
A_{c}(t)=A_{0}^{2}\sin (\omega t)/4\omega -A_{0}^{2}t/2.
\label{monochcorecpotentialfield}
\end{equation}

If we consider the limit $I_{p}\rightarrow 0$, i.e. the electron reaches
the continuum with zero kinematical momentum, the drift momentum of Eq.~(\ref{saddle1}) written in terms of the ponderomotive energy yields
\begin{equation}
\mathbf{k}=-2\sqrt{U_{p}}\cos (\omega t^{\prime })+\epsilon\left(\frac{U_{p}}{\omega }\sin (\omega t^{\prime })-2U_{p}t^{\prime }\right).
\label{newdriftmomentum}
\end{equation}

\subsubsection{In terms of electron energy}

\label{gmorbitsenergy}

We consider the final momentum $\mathbf{p}$ is parallel to the laser field
and write it in terms of kinetic energy to illustrate the simple-man model
for ATI. We solve the saddle point equations defined in Eqs.~(\ref%
{saddle1dir})-(\ref{saddle3}) in terms of the ionization $t^{\prime}$ and
rescattering $t$ times, respectively. For more close analysis, we restrict ourselves just
to the solutions of the first 3 cycles of our defined monochromatic field,
as shown in Fig.~\ref{pulse}. Classically, it is most probable that the electron
ionizes at the electric field maxima and returns to its parents ion at the
electric field crossings, i.e. when $E(t)=0$. In Fig.~\ref{pulse}(a) we depict the cases when the
electron returns to the core at a time about $\pi$ ($t=\pi/\omega\approx 55$ in a.u) while it tunnels at the
field maxima at times about $-\pi/2$ ($t\approx -28$ a.u.), $-3\pi/2$ ($t\approx -83$ a.u.), $-5\pi/2$ ($t\approx -138$ a.u.), and $-7\pi/2$ ($t\approx -193$ a.u.). 
A similar analysis was carried out in Ref.~\cite{Kopold1999} for an homogeneous field. Figure \ref{pulse}(b) shows the cases when the 
electron returns to the core at a time about $-\pi$ ($t\approx -55$ a.u.) while it tunnels at the field maxima at times about $\pi/2$ ($t\approx 28$ a.u.), $3\pi/2$ ($t\approx 83$ a.u.), $5\pi/2$ ($t\approx 138$ a.u.), and $7\pi/2$ ($t\approx 193$ a.u.).

\begin{figure}[th]
\begin{center}
\includegraphics[width=9cm]{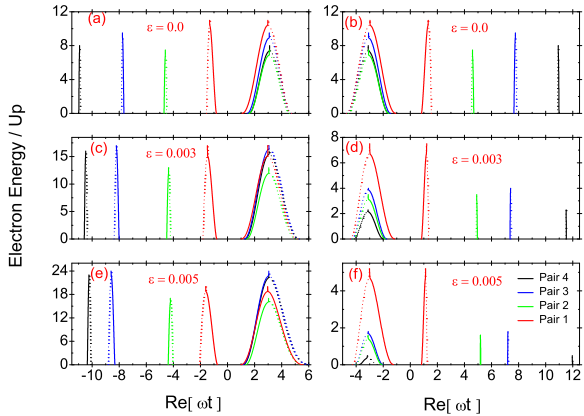}
\end{center}
\caption{(color online) Energy of the rescattered electron as a function of
the real part of the release time $t^{\prime }$ and the rescattered time $t$
of the electron for all given pairs in Fig.~(\protect\ref{pulse}). We
consider an hydrogen atom, for which the ground-state energy is $I_{p}=-0.5$
a.u., in a linearly polarized, monochromatic field of frequency $\omega=0.057$ a.u. (corresponding to a wavelength $\lambda=800$) and intensity $I=3\times10^{14} \mathrm{W/cm^{2}}$.
Panels (a) and (b) correspond to $\epsilon=0$, while panels (c)
and (d) show $\epsilon=0.003$ and panels (e) and (f) depict $\epsilon=0.005$. The left and right columns show the solutions
corresponding to the pairs given in Fig.~\ref{pulse}(a) and Fig.~\ref{pulse}(b), respectively. The dashed and solid lines correspond to the
long and the short orbits, respectively.}
\label{ATI1}
\end{figure}

In Fig.~\ref{ATI1}, we plot the energy of the rescattered electron as a
function of the real part of the ionization $t^{\prime }$ and rescattering $t$ times for the case with 
$\epsilon=0$ (panels (a) and (b)), $\epsilon=0.003$ (panels (c) and (d)) and $\epsilon=0.005$ (panels (e)
and (f)). The panels in the left and right columns show the solutions
correspond to the pairs given in Fig.~\ref{pulse}(a) and Fig.~\ref{pulse}(b),
respectively. 

For the homogeneous case, the ionization and rescattering times corresponding
to each cycle are identical, i.e. for the pairs when electron leaves at
about $-(n+1/2)\pi$ and returns at about $\pi$ (panel (a)) and for the pairs
when it leaves at about $(n+1/2)\pi$ and returns at about $-\pi$ (panel (b)). However, this is not true for the spatial inhomogeneous case. In here, for the case when electron leaves at about $-(n+1/2)\pi$ and returns at about $\pi$ (panels (c) and (e)) the energy cutoff is larger, while for the case when electron leaves at about $(n+1/2)\pi$ and returns at about $-\pi$ (panels (d) and (f)) the
energy cutoff is much smaller. In general, as a function of inhomogeneity of the field $\epsilon$, for the former case the energy cutoff will be
extended, while for the latter case the energy cutoff will move toward lower energy values. For the homogeneous case, the
shortest pair ($Pair_{1}$) has the largest cutoff, while for inhomogeneous case,
the $\epsilon$ will determine the pair with largest cutoff. For instance,
for $\epsilon=0.005$ (panel (a)) $Pair_4$ has larger cutoff than $Pair_1$.

\begin{figure}[th]
\begin{center}
\includegraphics[width=9cm]{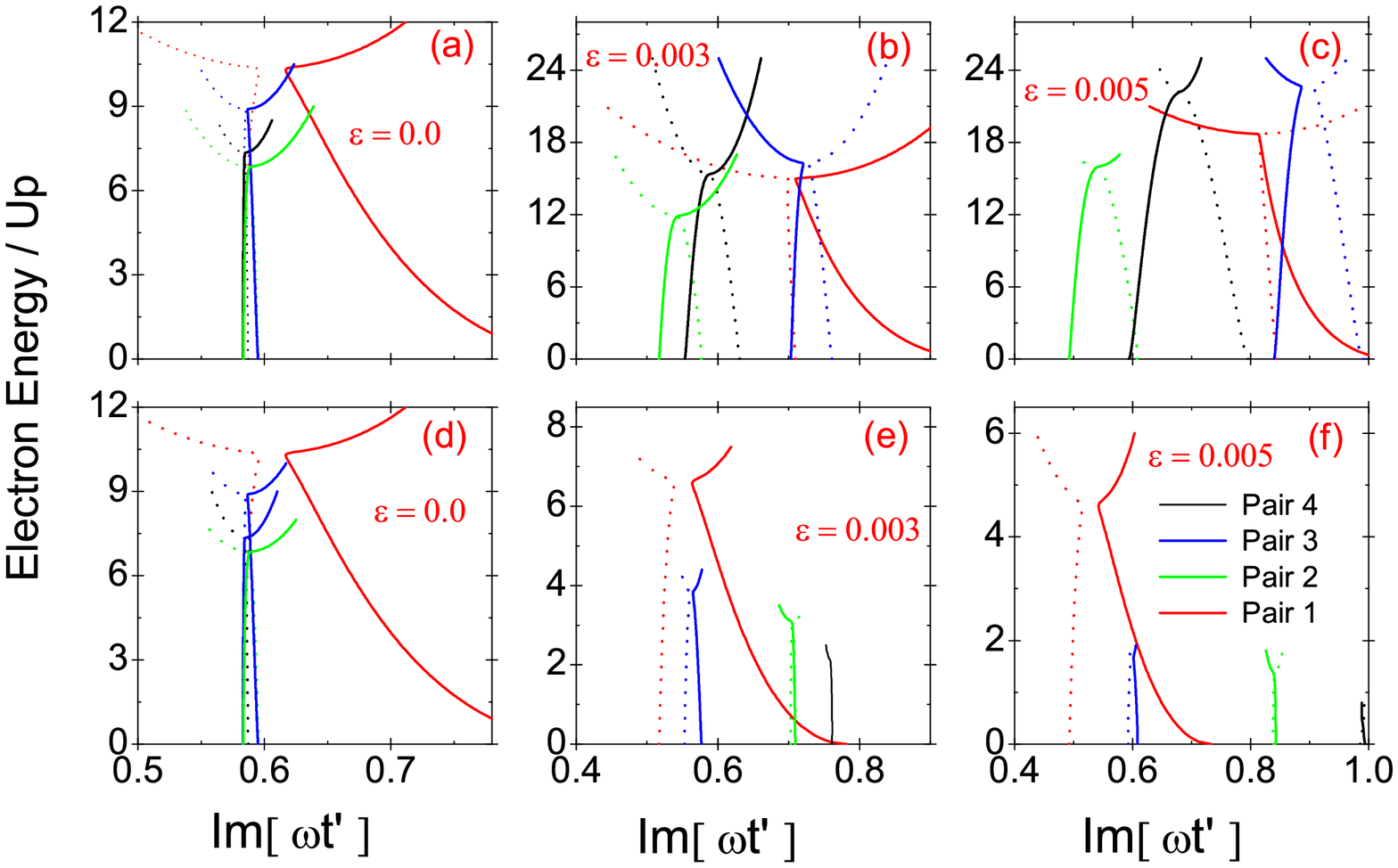}
\end{center}
\caption{(color online) Energy of the rescattered electron as a function of
the imaginary part of the release time $t^{\prime }$ of the electron for the
same parameters as in Fig.~\ref{ATI1} and for all given pairs of Fig.~\ref{pulse}. Panels (a) and (d) correspond to $\epsilon=0$, 
while panels (b) and (e) show $\epsilon=0.003$ and panels
(c) and (f) depict $\epsilon=0.005$. The upper and lower panels
show the solutions correspond to the pairs given in Fig.~\ref{pulse}(a) and Fig.~\ref{pulse}(b), respectively. The dashed and solid lines
correspond to the long and the short orbits, respectively.}
\label{ATI2}
\end{figure}

Figure \ref{ATI2} depicts the energy of the rescattered electron as a function
of the imaginary part of the release time $t^{\prime }$ of the electron, for
the case with $\epsilon=0$ (panels (a) and (d)), $\epsilon=0.003$
(panels (b) and (e)) and $\epsilon=0.005$ (panels (c) and (f)). The upper
and lower panels show the solutions correspond to the pairs given in Fig.~\ref{pulse}(a) and Fig.~\ref{pulse}(b), respectively.

For the homogeneous case, the imaginary part of the release time $t^{\prime }$
corresponding to each cycle are identical, i.e. for the pairs when electron
leaves at about $-(n+1/2)\pi$ and returns at about $\pi$ (panel(a)) and for
the pairs when it leaves at about $(n+1/2)\pi$ and returns at about $-\pi$
(panel(d)). However, this is not anymore valid for the inhomogeneous cases. The
inhomogeneity strength of the field $\epsilon$ strongly affects the imaginary
part of the release time $t^{\prime }$, and as a result, the ionization rate of
the electron.

\begin{figure}[th]
\begin{center}
\includegraphics[width=9cm]{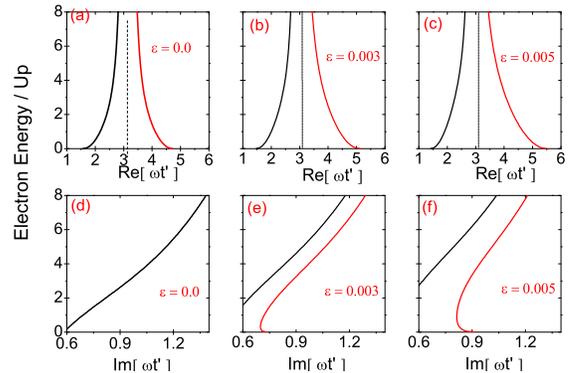}
\end{center}
\caption{(color online) Energy of the direct electron as a function of the
real and imaginary part of the release time $t^{\prime }$ for the same
parameters as in Fig.~\ref{ATI1}. Panels (a) and (d) correspond to $\epsilon=0$, while panels (b) and (e) show $\epsilon=0.003$ 
and panels (c) and (f) depict $\epsilon=0.005$. The upper
and lower panels show the real and imaginary time of the ionization,
respectively. The black and red lines show when electron leaves the atom at $\pi/2$ and $3\pi/2$, respectively.}
\label{ATI3}
\end{figure}

In Fig.~\ref{ATI3}, we present the energy of the direct electron as a function
of the real and imaginary part of the release time $t^{\prime}$. For the
homogeneous case, the imaginary part of the ionization time is the same for
the case when electron leaves at $\pi /2$ and $3\pi /2$ as shown in
panel (d). For this case the real time of trajectories are symmetric at $\pi $
(panel (d)). On the other hand, for the inhomogeneous case, the imaginary
time of the ionization is different for the case when electron leaves at $\pi/2$ and $3\pi/2$ as shown in panels (e) and (f). The deviation becomes
larger as function of $\epsilon$. In addition, the real part of the
saddle point equations when the electron tunnels at $\pi /2$ and $3\pi /2$
become less symmetric at around $\pi $ (panels (b) and (c)).

\subsubsection{In terms of electron momentum}

\label{gmorbitsmomentum}

We now investigate the solutions of the saddle point equations defined in
Eqs.~(\ref{saddle1dir})- (\ref{saddle3}) in momentum space. In order to
analyze this, we rewrite Eqs.~(\ref{saddle1dir}) and (\ref{saddle3})
in terms of the electron momentum components parallel $p_{\parallel }$ and perpendicular $p_{\perp }$ to the laser-field polarization,
\begin{equation}
\left[ p_{\parallel }+\mathbf{A}(t^{\prime })\right] ^{2}+4\epsilon \lambda
(t^{\prime })=-2I_{p}-p_{\perp }^{2}-4\epsilon A_{c}(t^{\prime
})p_{\parallel }  \label{saddle1dirmom}
\end{equation}
with
\begin{equation}
\lambda (t^{\prime })=A_{c}(t^{\prime })[p_{\parallel }+\mathbf{A}(t^{\prime
})]+\epsilon A_{c}^{2}(t^{\prime }),  \label{saddle4dirmom}
\end{equation}
and
\begin{equation}
\frac{\left[ p_{\parallel }+\mathbf{A}(t)\right] ^{2}}{2}=\frac{\left[
\mathbf{k}+\mathbf{A}(t)\right] ^{2}}{2}+2\epsilon A_{c}(t)[\mathbf{k}%
-p_{\parallel }-p_{\perp }]-p_{\perp }^{2}.  \label{saddle3mom}
\end{equation}

We now solve the saddle point equations in terms of the ionization $t^{\prime }$ and rescattered $t$ times for giving parallel $p_{\parallel }$ and perpendicular (transverse) $p_{\perp }$ momenta. We focus on the solutions of the shortest orbits for returning
electron considering our defined monochromatic field in Sec.~\ref{gmorbits}.
The remaining sets of the orbits are strongly suppressed due to the
wave-packet spreading. In particular, we consider the cases when the
electron tunnels around $\pi /2$ and rescatters at $2\pi $ and when
the electron tunnels at $3\pi /2$ and rescatters at $3\pi $. We refer to them
as the first shortest pair ($Pair_{s1}$) and the second shortest pair ($Pair_{s2}$), respectively.

\begin{figure}[tbp]
\begin{center}
\includegraphics[width=9cm]{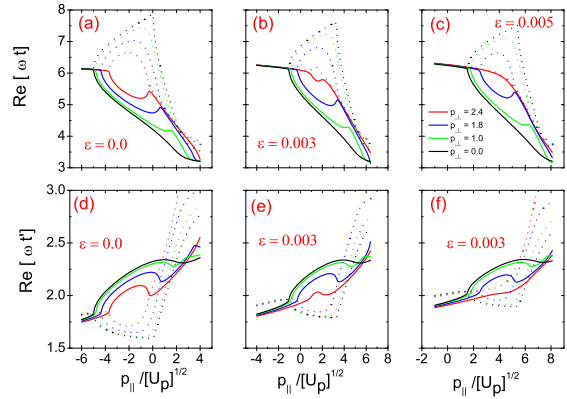}
\end{center}
\caption{(color online) Real part of the tunneling $t'$ and rescattering $t$ times
for the electron, as a function of its parallel momentum $p_{\parallel }$,
when the electron tunnels around $\pi /2$ and rescatters at about $2\pi $. We consider an hydrogen atom, for which the ground-state
energy is $I_{p}=0.5$ a.u., in a linearly polarized, monochromatic field of
frequency $\protect\omega=0.057$ a.u. and intensity $I=3\times10^{14}
\mathrm{W/cm^{2}}$. Panels (a) and (d) correspond to $\epsilon=0$, while panels (b) and (e) show $\epsilon=0.003$ and panels (c)
and (f) depict $\epsilon=0.005$. The upper and lower panels show
the rescattering Re[$t$] and the tunneling Re[$t^{\prime}$] times of the
electron, respectively. The dashed and solid lines correspond to the long
and the short orbits, respectively. }
\label{ATI1mom}
\end{figure}

Figure \ref{ATI1mom} demonstrates the real part of the solution of saddle
point equations of the rescattered electron, as function of the electron
momentum $p_{\parallel }$, for several transverse momenta, for the $Pair_{s1} $. The upper and lower panels correspond to the rescattering and
ionization times, respectively. The longer (dashed line) and the shorter (solid line)
orbits, along which the electron returns, can be identified, since the real
parts of $t^{\prime }$ and $t$ are associated to the classical trajectories
of an electron in a laser field. The long and short orbits coalesce for the
minimum and maximum momenta $p_{\parallel }$ for which the rescattering
process described by saddle-point Eq.~(\ref{saddle3mom}) has a classical
counterpart. Beyond these momenta we are in the classically not allowed
region, in which the yield decays exponentially.

For the homogeneous case, the classically allowed region is centered around $-2\sqrt{U_{p}}$ (Fig.~\ref{ATI1mom}(a)). This center is determined by the
most probable momentum the electron may have at the time of the
rescattering. In here, for vanishing transverse momentum $p_{\perp }$, the
cutoffs are near $-5\sqrt{U_{p}}$ and $2\sqrt{U_{p}}$ . The classically
allowed region shrinks and it results more localized around the center by increasing $p_{\perp }$. For larger $p_{\perp }$ this region completely disappears.

On the other hand, for the inhomogeneous case, the center of the classically allowed region is
not anymore at around $-2\sqrt{U_{p}}$ and it shifts towards different
momentum values. Figure \ref{ATI1mom} demonstrates how the
center and the extension of the classically allowed region strongly depend
on the value of $\epsilon$. For the case $\epsilon =0.003$, this center is around $1.5\sqrt{U_{p}}$
and the cutoffs are near $-\sqrt{U_{p}}$ and $5\sqrt{U_{p}}$ for $p_{\perp
}=0$ (Fig.~\ref{ATI1mom}(b)). Furthermore, for the case $\epsilon =0.005$, this center
is around $5\sqrt{U_{p}}$ and the cutoffs are near $\sqrt{U_{p}}$ and $6\sqrt{U_{p}}$ for vanishing transverse momentum $p_{\perp }$ 
(Fig. \ref{ATI1mom}(c)). As overall, the classical allowed region decreases by
increasing the inhomogeneity factor of the field $\epsilon$. In addition, this region shrinks much faster and more localized around the
center as $\epsilon$ increases.  

\begin{figure}[tbp]
\begin{center}
\includegraphics[width=9cm]{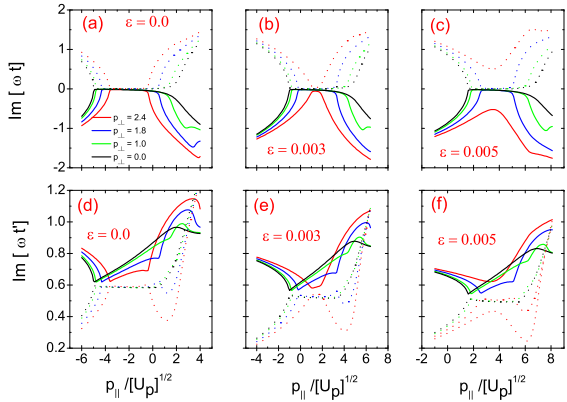}
\end{center}
\caption{(color online) Imaginary part of the tunneling $t'$ and rescattering $t$
times for the electron, as a function of its parallel momentum $p_{\parallel}$ for the same parameters as in Fig.~\ref{ATI1mom}. Panels (a) and
(d) correspond to $\epsilon=0$, while panels (b) and (e) show $\epsilon=0.003$ and panels (c) and (f) depict $\epsilon%
=0.005$. The upper and lower panels show the rescattering Im[$t^{\prime }$]
and the ionization Im[$t$] times of the electron, respectively. The dashed
and solid lines correspond to the long and the short orbits, respectively. }
\label{ATI2mom}
\end{figure}

Figure \ref{ATI2mom} demonstrates the imaginary parts of the ionization $t^{\prime }$ (upper
panels) and the rescattering $t$ (lower panels) times, for the $Pair_{s1}$. These solutions confirm the classical allowed
and not allowed regions interpretation. Indeed, they demonstrate that the
imaginary part of the rescattering times vanishes between the momenta for
which the real part of the rescattering times coalesces. Beyond this region
the imaginary $t$ increases abruptly. This is a very clear indication that
for the former and the latter regions the rescattering is classically allowed
and forbidden, respectively. In addition, the Im[$t^{\prime }$] exhibits a
minimum near the center of the classically allowed region, even if there is
no classically allowed region. It means that rescattering is most probable
for this specific momentum. On the other hand, the imaginary part of the
ionization time Im[$t^{\prime }$] of the electron is always non-vanished,
due to the fact that the tunneling process has no classically counterpart.
Furthermore, Im[$t^{\prime }$] gives some indication on how the width of the
potential barrier that the electron must overcome in the order to reach the
continuum, varies with regards to $\epsilon$. In general, this potential
barrier becomes wider as Im[$t^{\prime }$] increases. In here, the Im[$t^{\prime}$]
decreases slightly as a function of $\epsilon$.

\begin{figure}[tbp]
\begin{center}
\includegraphics[width=9cm]{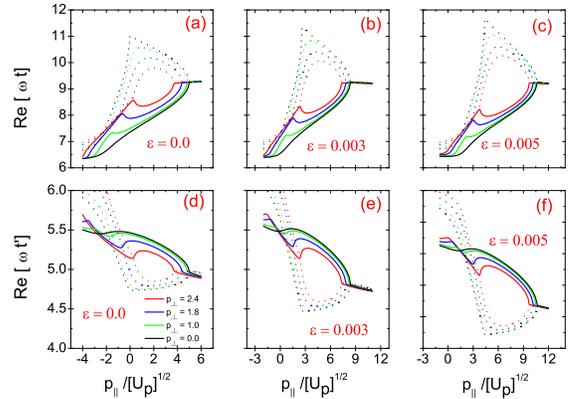}
\end{center}
\caption{(color online) Real part of the tunneling $t'$ and rescattering $t$ times
for the electron, as a function of its parallel momentum $p_{\parallel }$,
when the electron tunnels out at around $3\pi /2$ and rescatters at about $3\pi $, for the same parameters as in Fig.~\ref{ATI1mom}.
Panels (a) and (d) correspond to $\epsilon=0$, while panels (b)
and (e) show $\epsilon=0.003$ and panels (c) and (f) depict $\epsilon=0.005$. The upper and lower panels show the ionization Re[$t$] and the rescattering Re[$t^{\prime}$] times of the electron,
respectively. The dashed and solid lines correspond to the long and the
short orbits, respectively.}
\label{ATI3mom}
\end{figure}

Figure \ref{ATI3mom} demonstrates the real part of the solution of saddle
point equations of the rescattered electron, as function of the electron
momentum $p_{\parallel }$, for several transverse momenta and for the $Pair_{s2} $. 
The upper and lower panels correspond to the rescattering and ionization times.

For the homogeneous case, the classically allowed region is now centered at a
positive parallel momentum $p_{\parallel }$ (Fig.~\ref{ATI3mom}(a) and (b)) but it remains the same as
the $Pair_{s1}$ (Fig.~\ref{ATI1mom}(a) and (b)). On the contrary, for the inhomogeneous case, the classical allowed region is not the same as
the $Pair_{s1}$. Indeed, the classical allowed region is now much larger than
the case of the the $Pair_{s1}$. In contrast to the previous case, the
classical allowed region increases and more slowly shrinks and localizes
around the center by increasing the inhomogeneity factor of the field $\epsilon $. For instance, for the case $\epsilon =0.003$, the center of the
classically allowed region is around $4\sqrt{U_{p}}$ and the cutoffs are
near $0$ and $8.5\sqrt{U_{p}}$ for $p_{\perp }=0$ (Fig.~\ref{ATI3mom}(b)).
In addition, for the case $\epsilon =0.005$, this center is around $6\sqrt{U_{p}}$ and
the cutoffs are near $1.5\sqrt{U_{p}}$ and $11\sqrt{U_{p}}$ for vanishing
transverse momentum $p_{\perp }$ (Fig.~\ref{ATI1mom}(c)).

\begin{figure}[tbp]
\begin{center}
\includegraphics[width=9cm]{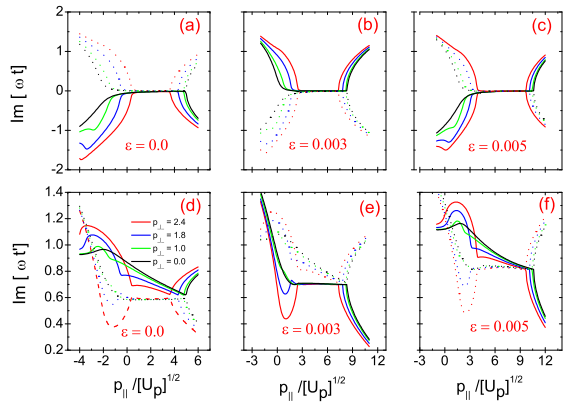}
\end{center}
\caption{(color online) Imaginary part of the tunneling $t'$ and rescattering $t$
times for the electron, as a function of its parallel momentum $p_{\parallel
} $ for the same parameters as in Fig.~\ref{ATI3mom}. Panels (a) and
(d) correspond to $\epsilon=0$, while panels (b) and (e) show $\epsilon=0.003$ and panels (c) and (f) depict 
$\epsilon=0.005$. The upper and lower panels show the rescattering Im[$t$]
and the ionization Im[$t^{\prime}$] times of the electron, respectively. The dashed
and solid lines correspond to the long and the short orbits, respectively.}
\label{ATI4mom}
\end{figure}

Figure \ref{ATI4mom} demonstrates the imaginary part of times $t^{\prime }$ (upper
panels) and $t$ (lower panels) of the $Pair_{s2}$. These solutions confirm the classical allowed
and not allowed regions interpretation of Fig.~\ref{ATI3mom}. In contrast to the previous case, the Im[$t$] decreases slightly as a function of $\epsilon$, indicating the widening of the potential barrier through which the electron must tunnel out.

\begin{figure}[h]
\begin{center}
\includegraphics[width=9cm]{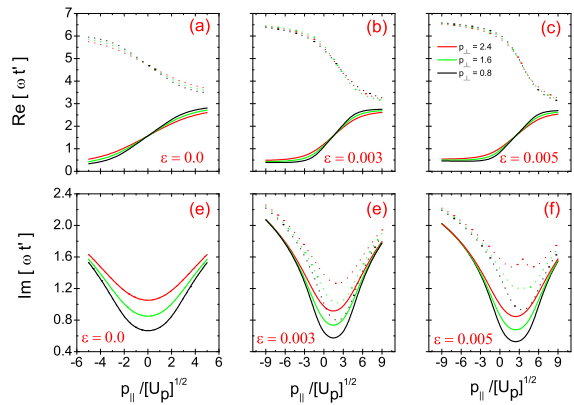}
\end{center}
\caption{(color online) Tunneling time $t'$ of the direct electron as a function of its parallel momentum $p_{\parallel }$, for several
transverse momenta $p_{\perp}$. Panels (a) and (d) correspond to $\epsilon=0$, while panels (b) and (e) show $\protect\epsilon=0.003$
and panels (c) and (f) depict $\protect\epsilon=0.005$. The upper and
lower panels give the real Re[$t^{\prime}$] and the imaginary Im[$t^{\prime}$] parts of such
times, respectively. The dashed and solid lines correspond to the long and
the short orbits, respectively.}
\label{ATI5mom}
\end{figure}

In Figure \ref{ATI5mom}, we plot the real and imaginary parts of the
solution of the saddle point equation of the direct electron (Eq.~(\ref{saddle1dirmom})), as a function of the electron momentum $p_{\parallel }$ and for
several transverse momenta. In all cases, the imaginary parts of tunneling time $t^{\prime}$ exhibit a minimum at the peak-field
times $\omega t= \pi/2$. This is due to the fact that the effective
potential barrier through which the electron tunnels out is narrowest for these
times. On the other hand, as the transverse momentum $p_{\perp}$ becomes
larger the Im[$t^{\prime }$] increases. This is consistent with the fact
that the potential barrier widens in this case.

For the homogeneous case, the imaginary time Im [$t^{\prime }$] of the neighboring orbits are identical and
has a minimum at $p_{\parallel}=0$. In here, two neighboring orbits behave symmetrically
with respect to the laser electric field and a electron sees the same effective
potential barrier for both orbits. Furthermore, the electron most probably
will tunnel out with zero momentum. On the other hand, for the inhomogeneous
case, the two orbits do not behave symmetrically with respect to the laser electric field
and the Im [$t^{\prime }$] is different for each orbit. Indeed, this time is
larger when electron tunnels at $t= 2\pi$ (dashed line) comparing to the
case when electron tunnels at $t= \pi/2$ (solid line). In addition, for the
inhomogeneous case, the electron most probably reaches the continuum with
a non zero momentum. Instead, the inhomogeneity factor of the field $\epsilon$
determines the most probable momentum. For instance, for $\epsilon=0.003$
and $\epsilon=0.005$, the electron will tunnels most probably with $p_{\parallel}= 1.5\sqrt{U_{p}}$ and $p_{\parallel}= 3\sqrt{U_{p}}$,
respectively. It means by increasing $\epsilon$ the electron most
probably will tunnel out with larger momentum.

\subsection{Spectra}

\label{spectrasec}

In this section, we compute the spectra of the direct and the rescattered ATI with Eqs.~(\ref{ActionDir}) and (\ref{ActionResc}) using the saddle point method
developed in Sec.~\ref{saddleEqs}.

\begin{figure}[tbp]
\begin{center}
\includegraphics[width=9cm]{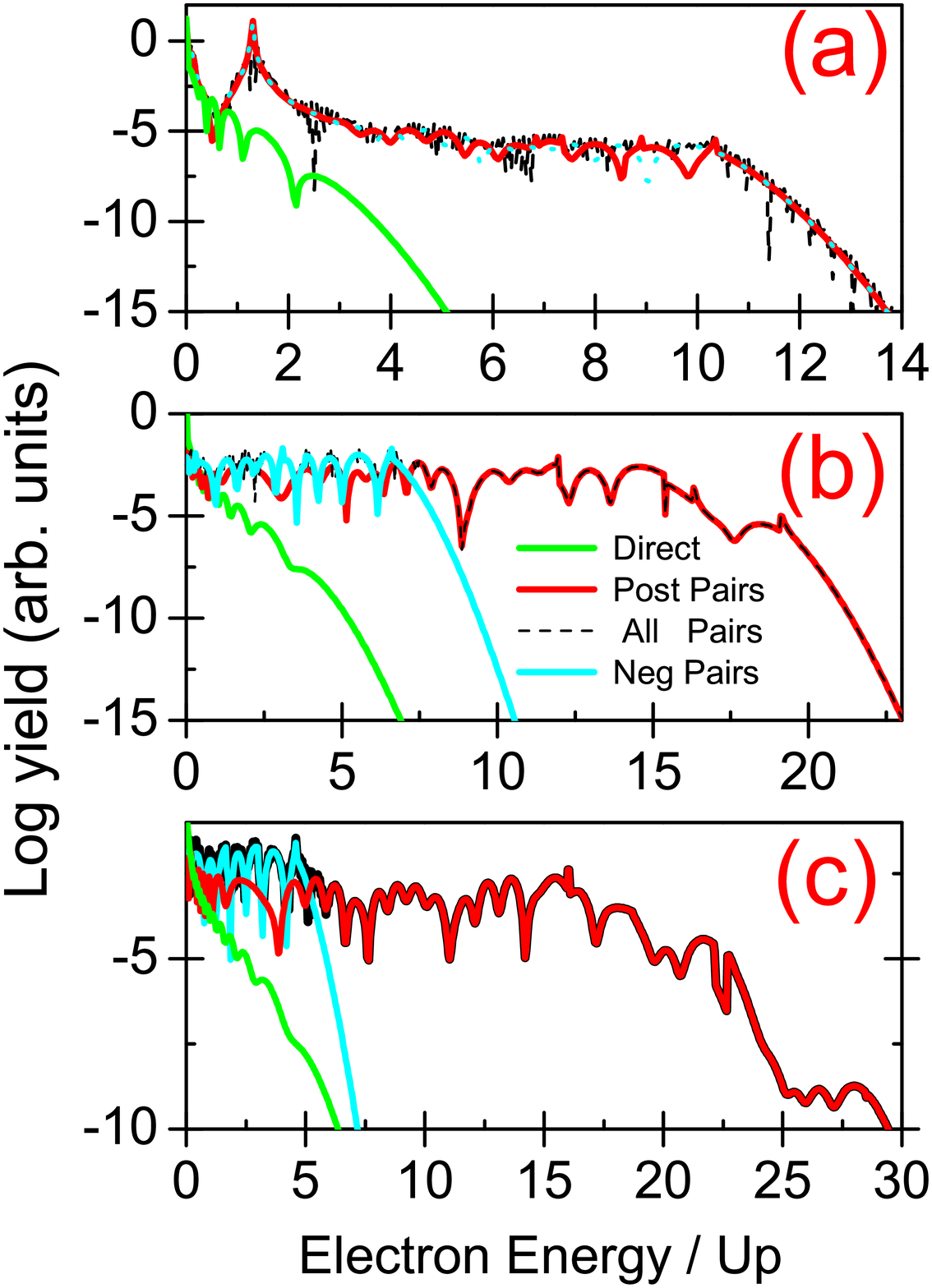}
\end{center}
\caption{(color online) ATI spectra for hydrogen atoms ($I_{p}=0.5$ a.u.)
interacting with a monochromatic field of frequency $\omega=0.057$
a.u. (corresponding to a laser wavelength $\lambda=800$ nm) and intensity $I=3\times10^{14} \mathrm{W/cm^{2}}$ for the case with $\epsilon=0.0$ (panel (a)), $\epsilon=0.003$ (panel
(b)) and $\epsilon=0.005$ (panel (c)). The green colored spectra
show the direct ATI and the rest show the rescattered ATI. The red and cyan
colored spectra show the contributions from pairs given in Fig.~\ref{pulse}(a) and Fig.~\ref{pulse}(b), respectively. The dashed black
colored spectra shows the total contributions from all pairs given in Fig.~\ref{pulse}.}
\label{spectra}
\end{figure}

Figure \ref{spectra} presents ATI spectra for hydrogen atoms ($I_{p}=-0.5$ a.u.)
interacting with a monochromatic field of frequency $\omega=0.057$
a.u. and intensity $I=3\times10^{14} \mathrm{W/cm^{2}}$ and for the cases with $\epsilon =0.0$ (homogeneous case), $\epsilon =0.003$, and $\epsilon =0.005$ as shown in panels (a), (b) and (c), respectively. In all panels the green line shows the direct ATI spectra. For the direct ATI, the cutoff is around the \textit{conventional} $2U_p$ value for the homogeneous
case and it extends to higher energy as a function of the inhomogeneity
factor of the field $\epsilon $. The cutoffs of the rescattered ATI are
in good agreement with the trajectories represented the previous Section (\ref{gmorbitsenergy}). For the inhomogeneous case, the pairs in Fig.~\ref{pulse}(a) lead to the largest cutoff (red lines) and the pairs in Fig.~\ref{pulse}(b) give lowest cutoff (cyan). However, the latter case will give the
final cutoff when we consider the total contributions, i.e. from all the pairs given in Fig.~\ref{pulse}, as shown in dashed black line. In the spectra with total contributions, the pairs from Fig.~\ref{pulse}(b) contribute to develop more interferences at the low energy part of the spectra. In conclusion, for the homogeneous case, the cutoff of the rescattered ATI is at the well known value of $10 U_p$ (Fig.~\ref{spectra}(a)), while these cutoffs are at around $17 U_p$ and $25 U_p$ for $\epsilon=0.003$ and $\epsilon=0.005$, respectively. This fact is indeed consistent with quantum mechanical simulations~\cite{ATICiappina2012}.

\subsection{Angular momentum distributions}

\label{momentumDist}

In this section, we calculate the electron momentum distributions as
a function of the electron momentum $p_{\parallel }$ and $p_{\perp }$ to the
laser field polarization and including both the direct and rescattered ionization
processes. We assume that that the prefactors Eqs.~(\ref{prefVp0}), (\ref{prefVk0})
and (\ref{prefVpk}) are constant, in order to remove any momentum bias that
may arise from such prefactors and having a clearer picture of how the
inhomogeneous strength influences such distributions.

\begin{figure}[h]
\begin{center}
\hspace*{-01.0cm}\includegraphics[width=10cm]{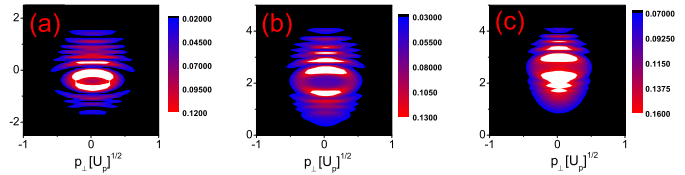}
\end{center}
\caption{(color online) Electron momentum distributions for direct ATI and for hydrogen atoms ($I_{p}=0.5$ a.u.) interacting with a
monochromatic field of frequency $\omega=0.057$ a.u. and intensity $I=3\times10^{14} \mathrm{W/cm^{2}}$. Panels (a), (b) and (c) show the case
with $\epsilon=0.0$, $\epsilon=0.003$ and $\epsilon=0.005$, respectively}
\label{DirMomDist}
\end{figure}

Figure \ref{DirMomDist} shows these calculations for the direct ATI.
For the homogeneous case (panel (a)), the distributions are centered at
around vanished $p_{\parallel }$ , while for the cases with $\epsilon=0.003$ and $\epsilon=0.005$ are at around $p_{\parallel }= 2%
\sqrt{U_{p}}$ and $p_{\parallel }= 2.5\sqrt{U_{p}}$. This fact is indeed consistent with
minima presented in lower panels of Fig.~\ref{ATI5mom}. For all cases, the overall
momentum space populated by the electron is the same and the interference
patterns are less pronounced as $\epsilon$ increases. Another interesting
feature is that the overall the yield increases with increasing of $\epsilon$, as shown in Fig.~\ref{DirMomDist}). 
This is due to the fact that the Im[$t$] decreases as a function of $\epsilon$, and as a result, the ionization rate increases.

\begin{figure}[h]
\begin{center}
\hspace*{-01.4cm}\includegraphics[width=10cm]{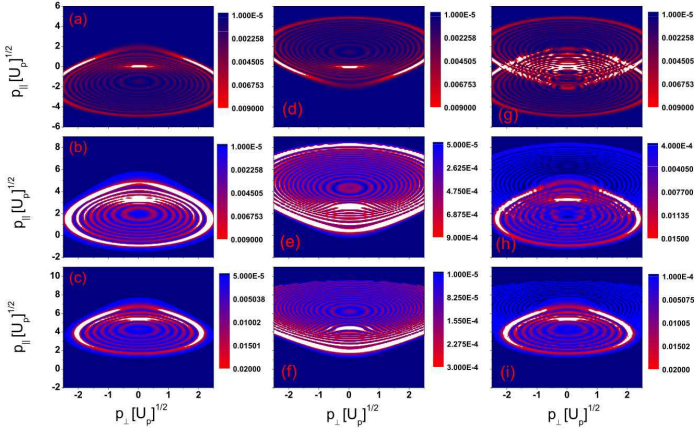}
\end{center}
\caption{(color online) Electron momentum distributions for rescattered ATI
and for hydrogen atoms ($I_{p}=0.5$ a.u.) interacting with a
monochromatic field of frequency $\omega=0.057$ a.u. and intensity $I=3\times10^{14} \mathrm{W/cm^{2}}$. The top, middle and bottom rows show the cases with $\epsilon=0$, $\epsilon=0.003$ and $\epsilon=0.005$, respectively. The left and middle columns
represent the case when the electron tunnels out around $\pi /2$ and
rescatters at about $2\protect\pi $ and when the electron tunnels at $3\pi /2$ and rescatters at $3\pi $, respectively, while the
right column shows when we have both cases.}
\label{ResMomDist}
\end{figure}

Finally, Fig.~\ref{ResMomDist}, shows the electron momentum distributions for the
rescattered ATI. The top row shows the homogeneous case, i.e. $\epsilon=0$, while the middle and bottom rows give the
inhomogeneous cases with $\epsilon=0.003$ and $\epsilon=0.005$,
respectively. The left and middle columns represent the cases for $Pair_{s1}$
and $Pair_{s2}$, respectively. The right column
depicts when both pairs are considered and added up coherently. For the $Pair_{s1}$, the electron momentum distributions are centered at 
around $p_{\parallel }= -2.0\sqrt{U_{p}}$, $p_{\parallel }= 2.0\sqrt{U_{p}}$ and $%
p_{\parallel }= 4.0\sqrt{U_{p}}$ for $\epsilon=0$, $\epsilon=0.003$,
and $\epsilon=0.005$, respectively. For the $Pair_{s2}$, these centers
are at around $p_{\parallel }= 2.0\sqrt{U_{p}}$, $p_{\parallel }= 4.0\sqrt{U_{p}}$ and $p_{\parallel }= 6.0\sqrt{U_{p}}$ for $\epsilon=0$, 
$\epsilon=0.003$, and $\epsilon=0.005$, respectively. For homogeneous
case, panels (a) and (d), both pairs have the same yield. On the other hand,
for the inhomogeneous cases, $Pair_{s2}$ has much smaller yield in comparison
to $Pair_{s1}$. We also calculate the electron momentum distributions by considering the contributions from both $Pair_{s1}$ and $Pair_{s2}$. These calculations are shown in panels (g), (h) and (i) for $\epsilon=0$, $\epsilon=0.003$,
and $\epsilon=0.005$, respectively. In here, for the inhomogeneous cases, $Pair_{s1}$ will dominate the total electron momentum contributions as it has higher yield.

\section{CONCLUSIONS}

\label{conclusion}

In this work, within the strong field approximation (SFA), we derive the transition amplitude for both direct and rescattered above-threshold-ionization (ATI) in an spatially inhomogeneous field. In particular, we show how the quantum orbits of the ATI (for both direct and rescattered processes) manifest themselves in spatially inhomogeneous fields. We show that in nonhomogeneous fields, the ionization and rescattering
times are not the same for the neighboring cycles, even for a monochromatic field. In here, the electron tunnels with different canonical momenta, which one leads to the lower energy cutoff and the other one develops a higher energy cutoff. We demonstrate that the energy cutoffs of the photoelectron spectra of the ATI extends to higher energies as a function of the inhomogeneity strength of the field. For both direct and rescattered processes, the center of the electron momentum distributions shift toward different momentum depending on the inhomogeneous factor and the cycle considered. Furthermore, the overall yield of the direct ionization slightly increases as function of inhomogeneous factor. For the rescattered case, this is true if the electron tunnels at around $\pi /2$ and rescattered at about $2\pi$. For the case when the electron tunnels at $3\pi /2$ and rescatters at $3\pi $, however, the yield decreases as function of the inhomogeneous strength.

\section{Acknowledgments}
We acknowledge the financial support of the MINCIN projects (FIS2008-00784
TOQATA and Consolider Ingenio 2010 QOIT) (M. F. C. and M.L.); ERC Advanced
Grant QUAGATUA, Alexander von Humboldt Foundation and Hamburg Theory Prize
(M. L.). We thank Samuel Markson for useful comments and suggestions.


\end{document}